\documentclass[12pt,showpacs,preprintnumbers,amsmath,amssymb,aps]{revtex4}
\newcommand{\step}{\vspace{.7em}}
\newcommand{\smallstep}{\vspace{.3em}}
\def\di{\displaystyle}
\def\tab{&\di}
\def\bg{\begin{eqnarray}\begin{array}{rcl}\displaystyle}
\def\eg{\end{array} &\di    &\di   \end{eqnarray}}
\def\bm#1{\begin{eqnarray}\begin{array}{#1}\di}
\def\bmo#1{\begin{eqnarray*}\begin{array}{#1}\di}
\def\bml#1#2{\begin{eqnarray}\begin{array}{#1}\label{#2}\di}
\def\bgo{\begin{eqnarray*}\begin{array}{rcl}\displaystyle}
\def\ego{\end{array} &\di    &\di \nonumber  \end{eqnarray*}}

\def\btensor#1#2{\renew\left#1\begin{array}{#2}\di}
\def\brtensor#1#2#3{\ren#3\left#1\begin{array}{#2}}
\def\botensor#1#2{\renew\left#1\begin{array}{#2}}
\def\etensor#1{\end{array}\right#1}

\def\eq#1{(\ref{#1})}
\def\Eq#1{Eq.~(\ref{#1})}

\def\d{{d}}

\def\Tr{{\rm Tr}}

\def\id{{\bf 1}}

\def\ov{\over}
\def\s0#1#2{\mbox{\small{$ \frac{#1}{#2} $}}}
\def\0#1#2{\frac{#1}{#2}}

\def\bQ{{Q}}

\def\CF{{\mathcal F}}
\def\CG{{\mathcal G}}

\def\CP{{\mathcal P}}

\def\CW{{\mathcal W}}

\newcommand{\GF}{{\frak F}}

\newcommand{\GG}{{\frak G}}

\def\ren#1{\renewcommand{\arraystretch}{#1}}

\def\renew{\renewcommand{\arraystretch}{1}}

\begin{document}

\title{Geometrical effective action and Wilsonian flows}

\vspace{1.5 true cm}

\author{
Jan M. Pawlowski}
\email{jmp@theorie3.physik.uni-erlangen.de}

\affiliation{\mbox{\it  
Institut f\"ur Theoretische Physik III, Universit\"at Erlangen,}\\ 
Staudtstra\ss e 7, D-91054 Erlangen, Germany}
\preprint{FAU-TP3-03-10}

\thispagestyle{empty}

\abstract{A gauge invariant flow equation is derived by applying a
  Wilsonian momentum cut-off to gauge invariant field variables. The
  construction makes use of the {\mbox geometrical} effective action
  for gauge theories in the Vilkovisky-DeWitt framework. The {\mbox
    approach} leads to modified Nielsen identities that pose
  non-trivial constraints on consistent truncations. We also evaluate
  the relation of the present approach to gauge fixed formulations as
  well as discussing possible applications.  }}
 \maketitle

\pagestyle{plain}
\setcounter{page}{1}
\section{Introduction}  

In the past decade the Exact Renormalisation Group 
\cite{Wilson:1973jj} in its modern
formulation \cite{Wetterich:yh} has lead to an impressive body of
results within scalar, fermionic as well as gauge theories, for
reviews see e.g.\ \cite{Litim:1998nf,Bagnuls:2000ae}.
For the scale dependent effective action $\Gamma_k$ it takes the
simple form
\begin{eqnarray}\label{flowintro}
\partial_t \Gamma_k[\phi]= \s012 \Tr\, \bigl(\Gamma_k^{(2)}[\phi]
+R\bigr)^{-1}\,\partial_t R, 
\end{eqnarray} 
where $R(p^2)$ introduces an infra-red regularisation to the theory with 
the infra-red cut-off scale $k$ and $t=\ln k$. The 
trace in \eq{flowintro} denotes a sum over momenta and internal
indices. Effectively, the momentum
integration in \eq{flowintro} only received contributions from an 
$R$-dependent momentum shell about $p^2=k^2$. This is responsible for the 
remarkable numerical stability of \eq{flowintro}. 
Furthermore, a
global optimisation procedure exists \cite{Litim:2000ci}, applicable
to general truncations. This optimisation hinges on the fact that the
regulator $R$ in \eq{flowintro} can be varied freely. The above
properties are at the root of reliability when using the flow equation
within some truncation to the full problem at hand. \smallstep 

In gauge theories, progress has been hampered by the intricacies of
implementing the underlying gauge symmetry in the presence of a
momentum cut-off. The key ingredient in \eq{flowintro} is the full
propagator, that is $\langle \phi(x)\phi(y)\rangle_{\rm 1PI}$. In a
gauge fixed formulation this object is not gauge invariant. Hence the
flow \eq{flowintro} breaks the gauge symmetry. The additional
technical difficulties are most clearly seen in a BRST formulation. A
purely algebraic constraint, the master equation, is turned into a one
loop equation similar to the flow equation, the modified
Slavnov-Taylor identities
\cite{Reuter:1993kw,Bonini:1993sj,Ellwanger:1994iz,D'Attanasio:1996jd,Litim:1998qi,Igarashi:2001ey}.
The loop term is proportional to $R\,\langle
\phi(x)\phi(y)\rangle_{\rm 1PI}$. The solution of these identities
within suitable truncations poses a considerable challenge.
\smallstep

Hence, quite some work has been devoted to 
gauge invariant formulations of Wilsonian flows
\cite{Reuter:1993kw,Litim:1998nf,Morris:1998kz} which also allow for practical
computations. In particular the background field approach to Wilsonian
flows \cite{Reuter:1993kw,Litim:2002ce} offers simplifications.  The
latter is of interest for the present formalism which is a natural
extension of the background field approach. We expect that
computational techniques and methods are similar. In the background
field approach the field is split into a sum of a background
configuration and a dynamical fluctuation field. This allows for the
definition of a gauge invariant, though gauge dependent, effective
action. Here gauge invariance is only an auxiliary symmetry and the
flow equation for the gauge invariant effective action is not closed
\cite{Pawlowski:2001df,Litim:2002xm}. The dynamical fluctuation still carries
the standard BRST symmetry which is modified in the presence of a
cut-off \cite{Litim:2002ce,Reuter:1997gx,Freire:2000bq,Bonini:2001xm}.
 The formalism allows for gauge
invariant truncations at the expense of loosing direct access to the
modified Slavnov-Taylor identities. Despite these deficiencies first
applications of the formalism to pure QCD
\cite{Reuter:1993kw,Pawlowski:1996ch,Reuter:1997gx,Liao:1995nm,Litim:2002ce,Gies:2002af}
as well as quantum gravity \cite{Reuter:1996cp} are rather promising.
Still, for convincingly treating the question of predictive power 
and for quantitative statements one has to employ more elaborated
truncations. For gauge invariant truncations, it is the gauge
dependence of physical results which signals their failure. The gauge
dependence of the effective action is encoded in the Nielsen identities 
\cite{Nielsen:fs} providing a benchmark test.  \smallstep

The geometrical approach was pioneered by Vilkovisky and DeWitt
\cite{Vilkovisky:st,DeWitt}. The crucial observation is that the
standard definition of the effective action even fails to maintain
reparametrisation invariance (of the fields in configuration space)
present in the classical action. This problem is resolved by coupling
geodesic normal fields to the current instead of the original fields.
Differences of these fields have a geometrical meaning in
contradistinction to differences of the original fields. In gauge
theories the dynamical geodesic normal fields are gauge invariant
leading to a gauge invariant and gauge independent effective action, for 
a brief review see \cite{Kunstatter:1991kw}.
As the propagator of the dynamical geodesic normal fields also is a
gauge invariant object, the problem of gauge invariance in
\eq{flowintro} is resolved. This interesting observation has been
recently put forward in \cite{Branchina:2003ek} for the sharp cut-off
case.  \smallstep

Here we present a derivation of a gauge invariant 
flow equation with general regulators for the geometrical effective
action. Gauge invariance and gauge independence is proven from the
flow equation.  The latter statement is misleading as the
construction depends on a background field serving as the base point
for the construction of geodesic normal fields.  This dependence is
encoded in the Nielsen identities, see e.g.\ 
\cite{Kunstatter:1991kw,Burgess:1987zi}. Modified Nielsen identities
in the presence of the cut-off are derived and their consequences are
evaluated. It turns out that these identities pose a
non-trivial consistency constraint on truncation schemes.  They
 are closely related to those posed by the standard
modified Slavnov-Taylor identities in a gauge fixed formulation. In
the view of these results we evaluate the limitations of the approach
in comparison to other formulations and outline possible practical
applications. For technicalities in both, the flow equation approach
and the Vilkovisky-DeWitt formalism, we have to refer the reader to
the literature. A self-contained introduction would go far beyond the
scope of this work, which is meant to present the key ideas, the main
technical steps as well as the important results and their interpretation.

\section{Geometry}

We use the formulation of the geometrical effective action as
presented in \cite{DeWitt:1998eq,Paris:en}.  Here we quote the results
relevant to us and essentially stick to the notation of
\cite{DeWitt:1998eq,leshouches}. For simplicity we only consider the
configuration space $\Phi$ of non-Abelian gauge fields $\varphi^i$ in
flat space-time; the present results straightforwardly generalise to
curved space-time and gravity. With DeWitt's condensed notation the
index $i=(x,\mu,\alpha)$ labels space-time $x$, Lorentz $\mu$, and 
gauge group $\CG$ indices $\alpha$. The gauge group is generated by vector
fields $\bQ_\alpha$ with Lie algebra
\begin{eqnarray}\label{lie}
[\bQ_\alpha,\ 
\bQ_\beta]= -c^\delta{}_{\alpha\beta} \bQ_\delta\,.
\end{eqnarray}
Here $c^\delta{}_{\alpha'\beta''}=f^\delta{}_{\alpha\beta}\,
\delta(x,x')\,\delta(x,x'')$, in a less condensed notation.
$f^\delta{}_{\alpha\beta}$ are the structure constants of the
non-Abelian group $G$. The components of the vector $\bQ_\alpha$ in 
the coordinates specified by $\{\varphi^i\}$ are
just $Q^i{}_\alpha=\bQ_\alpha \varphi^i$, the standard covariant
derivatives with gauge field $\varphi$.  Physics takes place in the
space $\Phi/\CG$. The group invariant metric $\gamma$ on $\Phi$ is
given by the ultra-local extension of the group invariant metric on
$\CG$. Its components in the chart $\{\varphi^i\}$ are given by 
$\gamma_{ij}=\gamma(\s0{\delta}{\delta \varphi^j},
\s0{\delta}{\delta \varphi^i})$. 
In case of a non-Abelian gauge group this
extension is unique and flat: $\gamma_\alpha{}^\mu{}_{\beta'}{}^{\nu'}=
\gamma_{\alpha\beta}\,\eta^{\mu\nu}\,\delta(x,x')$, where 
$\gamma_{\alpha\beta}$ is the Cartan-Killing metric of $G$. The 
horizontal projection $\Pi$ on $\Phi/\CG$ is defined by its components 
\begin{eqnarray}\label{horizontal} 
\Pi^i{}_j=\delta^i{}_j-Q^i{}_\alpha\omega^\alpha{}_j,  
\end{eqnarray}
where the indices refer to the coordinates $\varphi^{i}$ and 
$\omega^\alpha{}_j$ is  
\begin{eqnarray}\label{omega} 
{\omega}^\alpha_j=
\GG^{\alpha\beta} \bQ^j{}_\beta & 
\quad {\rm  with} &\GG^{-1}_{\alpha\beta}=
\gamma(\bQ_\alpha,\,\bQ_\beta). 
\end{eqnarray}
The connection 1-form 
$\omega^\alpha $ maps vertical vectors to the Lie algebra: 
$\omega^\alpha{}_i\, (\zeta^\beta Q^i{}_\beta)=\zeta^\alpha$. Hence 
$\Pi$ is just the projection onto
($\varphi$-) transversal modes: $Q^i{}_\alpha \Pi_i{}^j=0$, for
$\varphi=0$ it is the projection on transversal momentum
modes.\smallstep

The geometrical effective action hinges on a splitting of 
the degrees of freedom according to the projection $\Pi$. The
Vilkovisky connection $\Gamma_V$ (on the frame bundle $F\Phi$) relates
to a maximal disentanglement of the base space $F(\Phi/\CG)$ and the
gauge fibres. Indeed it is the pull-back of the Riemannian connection 
$\Gamma_g$ where $g$ is the restriction of $\Pi\cdot\gamma\cdot \Pi$ 
on $\Phi/\CG$. 
\begin{eqnarray}\label{Vcon}
\Gamma_V{}^i{}_{jk}=\Gamma_\gamma{}^i{}_{jk}-Q^i{}_{\alpha\cdot(j}\,
\omega^\alpha{}_{k)}-\s012\,\omega^\alpha{}_{(j}Q^i{}_{\alpha\cdot l}
Q^l{}_{\beta}\,\omega^\beta{}_{k)}.
\end{eqnarray}
In \eq{Vcon} $\Gamma_\gamma$ is the Riemannian connection on $F\Phi$ 
and subscripts ${}_{\cdot l}$ denote covariant derivatives based on
$\Gamma_\gamma$. The parenthesis in the subscripts indicate a
symmetrisation of the indices embraced implying that
$\Gamma_V{}^i{}_{jk}= \Gamma_V{}^i{}_{kj}$, the connection is torsion
free. The properties and consequences of $\Gamma_V$ are concisely
discussed in \cite{DeWitt:1998eq}. \smallstep 

The splitting of the degrees of freedom is achieved by introducing 
geodesic normal fields bases on $\Gamma_V$. First we choose a base
point $\varphi_*$ in $\Phi$. A gauge field $\varphi$ is described by
the coordinates of the tangent vector on the geodesic $\lambda(s)$ based on 
$\Gamma_V$: 
\begin{eqnarray}\label{gauss} 
\phi^i= (s-s_*){d(\varphi^i\circ \lambda)\ov d s}(s_*) 
\end{eqnarray} 
with $\lambda(s_*)=\varphi_*$ and $\lambda(s)=\varphi$ and $s$ is an
affine parameter of the geodesic $\lambda$. The construction entails
that the fields $(\Pi_*\phi)^A$ are invariant under a gauge
transformation of $\varphi$. Here we used the abbreviation
$\Pi_*=\Pi[\varphi_*]$ for the horizontal projection at the base
point. Only the fields $([\id-\Pi_*]\phi)^\alpha$ in the gauge fibre
at $\varphi_*$ change. Henceforth we reserve the
letters $i,j$ for coordinates specified by $\{\varphi^i\}$, 
letters $a,b$ for coordinates specified by $\{\phi^a\}$, 
letters $A,B$ for the gauge
invariant base space and $\alpha,\beta$ for the gauge fibre. \smallstep 

A gauge invariant 
functional $A[\varphi]$ satisfies $\bQ_\alpha A[\varphi]=0$. 
The above construction admits a covariant Taylor expansion 
of $A[\varphi]$: 
\begin{eqnarray}\label{covariant} 
A[\varphi_*,\phi]=\sum_{n=0}^\infty \0{1}{n!}\,A_{;(a_1\cdots 
a_n)}[\varphi_*] \phi^n.  
\end{eqnarray}
The semicolon denotes a covariant derivative based on the Vilkovisky
connection $\Gamma_V$. \Eq{covariant} implies that 
\begin{eqnarray}  
A^{(0,n)}_{,a_1\cdots 
a_n}[\varphi_*,0]= A_{; (a_1\cdots a_n)}[\varphi_*]\, ,
\label{derive} \end{eqnarray} 
where $A^{(n,m)}_{,a_1\cdots a_{n+m}}[\phi_1,\phi_2]$ denotes the $n$th
(covariant) derivative w.r.t.\ $\phi_1$ and the $m$th (covariant)
derivative w.r.t.\ $\phi_2$ of a general functional $A$. Henceforth we 
shall drop the superscript whenever it is redundant. It also can be shown that 
\begin{eqnarray}\label{,.}
A_{,a_1\cdots 
a_n}[\varphi_*,0]=A_{\cdot (b_1\cdots 
b_n)}[\varphi_*]\,\Pi_*{}^{b_1}{}_{a_1}\cdots \Pi_*{}^{b_n}{}_{a_n}
\end{eqnarray}
In the present case the metric $\gamma$ is flat leading to
 $\Gamma_\gamma=0$ and hence 
$A_{\cdot (b_1\cdots b_n)}[\varphi_*]=A_{, (b_1\cdots b_n)}[\varphi_*]$. 
With the help of \eq{covariant} and \eq{,.} one 
finally arrives at 
\begin{eqnarray}\label{invariant}
A[\varphi_*,\phi]
=\sum_{n=0}^\infty \0{1}{n!}\,A_{,a_1\cdots
a_n}[\varphi_*]\, (\Pi_*  \phi)^{a_1}\cdots (\Pi_* 
\phi)^{a_n}, 
\end{eqnarray}
that is, $A[\varphi]$ solely depends on the horizontally projected
geodesic normal coordinates $\Pi_*\phi$ as expected.

\section{Flow equation}

The classical Yang-Mills action only depends on the 
gauge invariant fields $\phi^A$ with
$S[\varphi]=S[\varphi_*,\Pi_*\phi]$.  The effective action can be
defined within a path integral only over the variables
$(\Pi_*\phi)^A=\phi^A$ with $\bQ_\alpha \phi^A=0$.  It owes gauge
invariance to the fact that its fields are scalars under gauge
transformations. It is gauge independent as no gauge is involved.
However, the fields $\phi^A$ are of formal use only so we introduce
gauge fixing terms as a technical tool at an intermediate level. It
simply allows us to extend the path integral to the full set of
geodesic normal fields $\phi^a$ instead of $ \Pi_*\phi$. The resulting
effective action is still gauge invariant and gauge independent.
For gauges linear in the field $\phi$ the gauge fixing term reads
\begin{eqnarray}\label{linear} 
S_{\rm gf}[\varphi_*,\phi]={1\ov 2\xi}
\kappa_{\alpha\beta}[\varphi_*] {P^\alpha}_a[\varphi_*]
{P^\beta}_b[\varphi_*] \phi^a \phi^b, 
\end{eqnarray} 
where the gauge fixing operator $P$ could depend on the base point
$\varphi_*$, but not necessarily. With minor modifications, 
the construction below also applies to more general gauge fixings
${P^\alpha}_a[\varphi_*,\phi] \phi^a$ including the standard choice
$Q_*^\alpha{}_i(\varphi-\varphi_*)^i$.  The action $S_{\rm gf}$ in
\eq{linear} simply provides a Gau\ss ian factor for the integration
over the gauge degrees of freedom ${P^\alpha}_a[\varphi_*]\phi^a$.
Moreover, one can always diagonalise such that
\begin{eqnarray}\label{diagonalP} P^{\alpha}{}_a\, \Pi_*{}^a{}_b =0.
\end{eqnarray} 
Thus we only have to consider gauge fixing terms \eq{linear}
with the property \eq{diagonalP}. Finally we introduce a momentum
cut-off to the theory by adding the term $\Delta S_k$ to the classical
action with
\begin{eqnarray}\label{cutoff} 
\Delta S_k[\varphi_*,\phi]= {1\ov 2}
R_{ab}[\varphi_*]\phi^a \phi^b.   
\end{eqnarray}
The regulator depends on an infra-red scale $k$ and supposedly
regularises the infra-red behaviour of the theory. We also allow for a
dependence on the base point $\varphi_*$ but not on $\phi$. For small
momenta $R$ should serve as a mass for the field $\phi$, whereas it
should decay for large momenta, rendering the ultra-violet regime
unaltered. With \eq{linear} and \eq{cutoff} we are led to the
effective action $\Gamma_k$:
\begin{eqnarray}
e^{\Gamma_k[\varphi_*,\bar\phi]}=\int 
[\d\phi^a]_{\rm ren}\,\mu[\varphi_*,\phi]\, J[\varphi_*,\phi]
\,e^{-S[\varphi_*,\phi]-S_{\rm gf}[ \varphi_*,\phi-\bar\phi]-
\Delta S_k[\varphi_*,\phi-\bar\phi]+
    \Gamma_{k,a}[\varphi_*,\bar\phi](\phi^a-\bar\phi^a)}. 
\label{pathk}
\end{eqnarray} 
The subscript ${}_{\rm ren}$ indicates that \eq{pathk} is defined as
discussed in \cite{Pawlowski:2001df}, i.e.\ the renormalisation of the
path integral is done such that the cut-off term $\Delta S_k$ carries
the only $k$-dependence.  The measure $\mu[\varphi_*,\phi]$ entails
the correct boundary conditions of the path integral and
ensures reparametrisation invariance of the measure
$[\d\phi^a]_{\rm ren}\,\mu[\varphi_*,\phi]=
[\d\varphi^i]_{\rm ren}\,\mu[\varphi]$. The $\phi^\alpha$ where only
introduced as auxiliary variables and the measure factorises:
$\mu=\mu_{\phi^A}\mu_{\phi^\alpha}$. For gravity and Yang-Mills theories 
we can identify 
\begin{eqnarray}\label{measure} 
\mu[\varphi_*,\phi]=\sqrt{\gamma[\varphi_*,\phi]},
\end{eqnarray} 
where $\gamma[\varphi_*,\phi]$ is the determinant of the metric
$\gamma_{ab}$. The general case is more involved \cite{leshouches}. 
The term $J[\varphi_*,\phi]$ is related to the Fadeev-Popov determinant
\begin{eqnarray}
J[\varphi_*,\phi] = \det
\GF[\varphi_*,\phi] \sqrt{\det \kappa} &\quad {\rm with} \quad & 
\GF^\alpha{}_\beta[\varphi_*,\phi] = P^\alpha{}_a[\varphi_*]\,
Q^a{}_\beta[\varphi_*,\phi] \label{J}
\end{eqnarray}
and $Q^a{}_\beta[\varphi_*,\phi]={\bQ}_\beta\phi^a$. It follows from
the definition \eq{pathk} that the effective action $\Gamma_k$ does
not depend on $\bar\phi^\alpha$ and is the standard geometrical
effective action for $k=0$. This is most easily seen within the
following specific choice of vertical geodesic fields: $d\varphi^i=
\varphi^i_{,A}\,d\phi^A+Q^i{}_\alpha\, d\phi^\alpha$, see e.g.\ 
\cite{Kunstatter:1991kw}. Then the determinant of the metric
decomposes into
\begin{eqnarray}\label{decompose}
\det\gamma_{ab}=\det g_{AB}\,\det \GG^{-1}_{\alpha\beta}, 
\end{eqnarray}
using $\gamma_{\alpha\beta}=\GG^{-1}_{\alpha\beta}$, see \eq{omega}.
The $\bar\phi^\alpha$-independence already ensures invariance of
$\Gamma_k$ under gauge transformations of $\bar\phi$ as they only
affect $\bar\phi^\alpha$. General regulators mix the propagation of
$\phi^A$ and $\phi^\alpha$. Even though this does not spoil gauge
invariance it spoils the disentanglement of $\phi^A$ and
$\phi^\alpha$. Still, the fields can be disentangled by appropriate
gauge transformations. As the $\phi^\alpha$ are only auxiliary
variables these problems can be avoided by only regularising the
propagation of $\phi^A$:
\begin{eqnarray}\label{onlyA}
R_{a_1 b_1}\,\Pi_*{}^{ a_1}{}_{a}\, \Pi_*{}^{b_1}{}_b=R_{ab}, 
\end{eqnarray}
in components: $R^{\alpha\beta}$=0. This naturally goes along with
\eq{diagonalP}. We shall discuss later how general regulators reduce
to this case.  Taking the derivative of \eq{pathk} w.r.t.\ the
logarithmic infra-red scale $t=\ln k$ we arrive at the flow equation
\begin{eqnarray}\label{flow} 
\dot\Gamma_k[\varphi_*,\bar\phi]= {1\ov 2}G^{ab}[\varphi_*,\bar\phi] \, 
\dot R_{ba}[\varphi_*],
\end{eqnarray} 
where $\dot F =\partial_t F$ 
and the full propagator $G$ is given by   
\begin{eqnarray}
G[\varphi_*,\bar\phi]= \bigl( 
    \Gamma_k^{(0,2)}+\s01\xi P\kappa P+R\bigr)^{-1}[\varphi_*,\bar\phi]   
 \label{prop} \end{eqnarray} 
with $(P\kappa P)_{ab}=\kappa_{\alpha\beta} {P^\alpha}_a {P^\beta}_b$
and $\Gamma_k^{(0,2)}$ is the second derivative w.r.t.\ $\bar\phi$ as
introduced below \eq{derive}. The one loop form of \eq{flow}
originates in the fact that the cut-off term \eq{cutoff} is only
quadratic in $\phi$ ($R$ is $\phi$-independent). Higher powers in
$\phi$ inevitably lead to higher loop terms in \eq{flow}
\cite{Litim:2002xm} spoiling its numerical and conceptual simplicity.
We emphasise that \eq{flow} is valid for general $R[\varphi_*]$, not
only for those satisfying \eq{onlyA}.

\section{Gauge invariance}

The flow equation \eq{flow} with the effective action
$\Gamma_\Lambda[\varphi_*,\bar\phi]$ at some initial scale $\Lambda$
serves as a non-perturbative definition of the path integral
\eq{pathk}. General properties of the theory can be
accessed via the flow equation and $\Gamma_\Lambda$ in a well-defined
setting, as \eq{flow} is both infra-red safe and ultra-violet finite. 
In particular one 
can study how the symmetries of the theory evolve with the flow. 
Accordingly, it should be possible to 
prove gauge invariance and gauge independence of
$\Gamma_k$ solely from \eq{flow} by assuming its validity for
$\Gamma_\Lambda$. We shall 
see that the effective action is invariant under the transformations
\begin{subequations}\label{gaugetrafos}
\begin{eqnarray}\label{gaugetrafo} 
{\bQ}_\alpha\bar\phi^a=Q^a{}_\alpha[\varphi_*,\bar\phi], \tab \quad \quad \tab 
{\bQ}_\alpha \varphi_* =0  
\end{eqnarray} 
and 
\begin{eqnarray}\label{bgaugetrafo} 
{\bQ_*^i}_\alpha \bar\phi^a{}_{;i}=Q_*^a{}_\alpha, &\quad \quad 
& {\bQ_*}_\alpha 
\varphi^i_* =Q_*^i{}_\alpha,  
\end{eqnarray} 
\end{subequations}
where $\bQ_*{}_\alpha=\bQ_\alpha[\varphi_*]$. The first equation in  
\eq{bgaugetrafo} follows from the properties of $\Gamma_V$, see e.g.\ \cite
{Vilkovisky:st,Fradkin:1983nw,Rebhan:1986wp}. 
The transformation \eq{gaugetrafo} relates to the quantum gauge
transformations in the background field formalism. It implies a gauge
transformation of $\bar\varphi=\varphi[\varphi_*,\bar\phi]$ at fixed
$\varphi_*$. The transformation \eq{bgaugetrafo} relates to the
background gauge transformation. It implies a gauge transformation of
$\varphi_*$ at fixed $\bar\varphi$. As $Q^a{}_\alpha$ can be expanded 
in $Q_*^a{}_\beta$, invariance of
$\Gamma_k[\varphi_*,\bar\phi]$ under \eq{gaugetrafos} can be expressed as 
\begin{subequations}\label{invariance} 
\begin{eqnarray}\label{invariance1} 
Q_*^a{}_\alpha \Gamma_{k,a}& =& 0,\\
Q_*^i{}_\alpha \Gamma_{k,i}& =& 0. 
\label{invariance2} \end{eqnarray}
\end{subequations}
Only \eq{invariance2} imposes a constraint on $\Gamma_k$.
\Eq{invariance1} just comprises the fact that the effective action
solely depends on $\Pi_*\phi$. We proceed with the  proof that the effective
action $\Gamma_k[\varphi_*,\bar\phi]$ satisfies \eq{invariance} and is
gauge independent for all $k$, if $\Gamma_\Lambda[\varphi_*,
\bar\phi]=0$ has these properties. 
This starting point is guaranteed for the classical
action (for $R\to \infty$), or alternatively for the full effective
action $\Gamma_0$ (for $R\to 0$), even though in the latter
case is rather a presupposition \cite{strictly}:

\begin{itemize}
\item[{\bf (i)}] {\it Invariance under gauge transformations of $\bar\phi$}
\smallstep

\noindent $\bQ_\alpha\Gamma_\Lambda=0$ entails that $\Gamma_\Lambda$ 
only depends on $\phi^A$ and $\varphi_*$. Then $\bQ_\alpha
\Gamma_\Lambda^{(0,2)}=0$ as $\Gamma_\Lambda^{(0,2)}$ is gauge
invariant ($\phi^\alpha$-independent). Hence the right hand side of
the flow equation is gauge invariant leading to $\bQ_\alpha\,
\partial_t\Gamma_k|_{k=\Lambda}=0$.  By recursion it follows that
$\bQ_\alpha\Gamma_k=0$ for all scales $k$, i.e.\ \eq{invariance1}.
The above argument is only based on the flow equation and is valid for
general $R$, not only those satisfying \eq{onlyA}.

\item[{\bf (ii)}] {\it Invariance under gauge transformations of 
$\varphi_*$}
\smallstep

\noindent $\bQ_*^i{}_\alpha
\Gamma_{\Lambda,i}=0$ entails that $
\bQ_*^i{}_\alpha\Gamma^{(1,2)}_{\Lambda,i}=0$ and with \eq{onlyA} 
it follows that $G,R,\dot R$ are
block-diagonal at $k=\Lambda$. In components this
reads $G^{\alpha A}=G^{A\alpha}=0$ and $R_{\alpha
  A}=R_{A\alpha}=R_{\alpha\beta}=0$.
 The gauge variation $Q_*^i{}_\alpha\dot\Gamma_{k,i}$
is given by
\begin{eqnarray}
Q_*^i{}_\alpha\dot\Gamma_{k,i}=  \s012 \left(
G^{ab}{}_{,i}\,\dot R_{ba}+G^{ab}\,\dot R_{ba,i}\right)\,
Q_*^i{}_\alpha.   
\label{step2}\end{eqnarray} 
\Eq{step2} vanishes if $Q_*^i{}_\alpha R_{ba,i}$ has no diagonal 
($AB$ and $\alpha\beta$) components and $Q_*{}_\alpha G^{ab}$ only has 
off-diagonal and $\alpha\beta$ components. We conclude that 
\eq{step2} vanishes only for 
$\Pi_*{}^c{}_a \Pi_*{}^d{}_b\, Q_*{}_\alpha R_{cd}=0$. Combined with  
\eq{onlyA} this reads in components 
\begin{eqnarray}\label{Rproperty}
\bQ_*{}_\alpha R_{AB}=0, &\qquad & R_{\alpha\beta}=0. 
\end{eqnarray} 
\Eq{Rproperty} entails $(\id-\Pi_*)^c{}_a (\id-\Pi_*)^d{}_b\,
Q_*{}_\alpha R_{ab}=0$. Consequently the
second term on the right hand side of \eq{step2} vanishes. With
\eq{diagonalP}, \eq{Rproperty} and 
$\bQ^i_*{}_\alpha\Gamma^{(1,2)}_{\Lambda,i}=0$ we conclude that 
$\Pi_*{}_c{}^a \Pi_*{}_d{}^b\, Q_*{}_\alpha G^{cd}=0$.  Then \eq{onlyA}
leads to a vanishing first term on the right hand side of \eq{step2}.
This results in $Q_*^i{}_\alpha \Gamma_{k,i}=0$ for all scales $k$,
i.e.\ \eq{invariance2}. 

 \item[{\bf (iii)}] {\it Gauge independence}\smallstep 
 
\noindent For regulators satisfying \eq{onlyA}, the 
effective action $\Gamma_k$ is independent of the gauge fixing $P\phi$
if it is gauge invariant \eq{invariance1}.  Assume it is valid at some
scale $\Lambda$. We consider variations $\delta_P$ of $P$ in the class
of gauge fixings \eq{diagonalP}. Applied to the flow equation \eq{flow} 
this leads to 
\begin{eqnarray}\label{Pindep}
\delta_P\dot\Gamma_\Lambda
=-\s0{1}{2\xi} (G  
\dot R G)_{ab} 
\,\delta_P (P\kappa P)^{ba}=0   
\end{eqnarray}
with
 $P(G \dot R G)=(G \dot R G) P=0$ following from \eq{diagonalP}, 
\eq{onlyA}. We also used that \eq{invariance1} implies that
 $P\Gamma_k^{(0,2)}=\Gamma_k^{(0,2)} P=0$.  
\end{itemize}

\noindent We close this section 
with discussing important consequences and implications of
{\bf(i)}-{\bf(iii)}:\smallstep 

The cut-off term supposedly regularises the
propagation of momentum modes.  According to \eq{Rproperty} the
related momenta (squared) $\CP^2$ have to be covariant under gauge
transformations of $\varphi_*$ as $\bQ_*{}_\alpha \CP^2_{AB}=0$. This
entails that regulators satisfying \eq{Rproperty} depend
on $\varphi_*$. \smallstep
 
In contradistinction to the geometrical effective 
action at 
$k=0$ the properties {\bf (i)} and {\bf (iii)} do not entail {\bf
  (ii)}. The former
only need \eq{onlyA} whereas for the latter \eq{Rproperty} is required. 
Naturally the question arises whether the constraint \eq{Rproperty} 
can be relaxed. To that end consider the regulators satisfying 
\begin{eqnarray}\label{preRproperty}
\bQ_*{}_\alpha R_{AB}=0, &\quad & R_{\alpha A}=0,\ \ R_{A\alpha}=0  
\end{eqnarray}
which in general have non-vanishing components $R_{\alpha\beta}\neq0$.  We get
invariance under \eq{invariance} and gauge independence for
$\bar\Gamma_k$, where $\bar\Gamma_k=\Gamma_k+C_k[\varphi_*]$ with
$C_k=-((1-\Pi_*)\ln (R+P\kappa P))_a{}^a$. Then it follows with
\eq{step2} that $ \bQ_*{}_\alpha\bar\Gamma_k=0$. The effective action
$\bar\Gamma_k$ is a shift of $\Gamma_k$ with a $\varphi_*$-dependent
constant and simply expresses the fact that $R_{\alpha\beta}$ can be
interpreted as an additional gauge fixing.  Hence, effectively we are
still dealing with \eq{Rproperty}. \smallstep

This consideration extends to general $R$: since $\phi^\alpha$ only  
appears quadratically in \eq{pathk} we always can diagonalise. 
Schematically the corresponding effective regulator    
$R_{\rm eff}$ is  
\begin{eqnarray}\label{effective}
R_{\rm eff}=\Pi_*\,R\left(1+(1-\Pi_*)\s0{1}{R+\s0{1}{\xi}P\kappa P}(1-\Pi_*)
R\right)\Pi_*. 
\end{eqnarray} 
In general effective regulators $R_{\rm eff}$ neither satisfy
\eq{Rproperty} nor \eq{preRproperty}. Still they satisfy \eq{onlyA}.
Consequently we only have lost invariance under gauge transformations
of $\varphi_*$ \eq{invariance2}. We emphasise that the effective
regulator $R_{\rm eff}$ is $\varphi_*$-dependent even if the original
regulator $R$ is not. Also, the effective gauge fixing follows as
below \eq{preRproperty}.  Even though all these generalisations can be
reduced to \eq{onlyA} the more general form proves useful in
applications.  \smallstep

In general gauge independence {\bf (iii)} cannot be proven if gauges
more general than \eq{diagonalP} are considered. This can be mapped to
the case of more general regulators considered above. As discussed there 
that general regulators relate to a modification of the
gauge fixing. We emphasise that by removing the $\phi^\alpha$ from the
theory all these subtleties disappear. The reason to keep them is that
they permit representations of the flow that facilitate practical
applications as well as general considerations. \smallstep

The findings above also allow us to device a gauge invariant  
effective action and its flow in terms of standard gauge fields
$\bar\varphi=\varphi[\varphi_*,\bar\phi]$. We simply rewrite the
effective action $\Gamma_k$ in terms of $\bar\varphi$:
$\hat\Gamma_k[\varphi_*,\bar\varphi]:=
\Gamma_k[\varphi_*,\bar\phi[\varphi_*,\bar\varphi]]$.  Invariance of
$\hat\Gamma_k$ under gauge transformations of $\bar\varphi$ follows
from invariance of $\Gamma_k$. The flow equation \eq{flow} is a
self-contained flow for $\hat\Gamma_k$ as $\partial_t
\hat\Gamma_k=\partial_t\Gamma_k$. The two-point function $\Gamma_k^{(0,2)}$ 
on the right hand side of \eq{flow} 
can be rewritten in terms of derivatives w.r.t.\ $\bar\varphi$ as 
\begin{eqnarray}\label{derivatives} 
{\delta\over \delta \bar\phi^a}= \left( \bar\phi_{;i}^a\right)^{-1}
{\delta\over \delta \bar\varphi^i}.
\end{eqnarray}
In \eq{derivatives} we have used that $\phi[\varphi_*,\bar\varphi]$ is
a scalar in the second variable (see e.g.\ \cite{DeWitt:1998eq,Paris:en}). 
For practical applications within truncation schemes one has to bear in mind 
that $\hat\Gamma_k$ is not the generating functional of 
1PI Green functions of $\bar\varphi$. It is equivalent to $\Gamma_k$
which generates 1PI Green functions of $\bar\phi$, see e.g.\ 
\cite{Burgess:1987zi}. \smallstep 

We close with the remark that the sharp cut-off
flow of \cite{Branchina:2003ek} is obtained by the following
procedure: take the sharp cut-off limit for $R$ in \eq{flow}, use
\eq{measure} and \eq{derivatives}. For the remaining notational
differences, see e.g.\ \cite{PaccettiCorreia:2002fw}. If invariance
under gauge transformations of $\varphi_*$ should be maintained {\bf
  (ii)}, $R$ has to satisfy \eq{Rproperty}. The sharp cut-off
limit has to be taken with respect to gauge invariant momenta squared.
It implies that $R$ and the sharp cut-off depend on $\varphi_*$ as it
cannot depend on $\bar\phi$.  This stays true for general regulators
as the corresponding effective regulators \eq{effective} are
$\varphi_*$-dependent.

\section{Modified Nielsen identities}

The above analysis makes it clear that the construction 
is not unique: $\Gamma_k[\varphi_*,\bar\phi]$ depends on the choice of
the base point $\varphi_*$, see also \cite{Branchina:2003ek}. 
We have traded gauge dependence and
modified Slavnov-Taylor identities for base point dependence. Indeed,
the latter should be interpreted as a gauge fixing. $\Gamma_k$ is the
generating functional for 1PI Greens functions of $\Pi_*\phi$. The
propagating degrees of freedom obey the Landau-DeWitt gauge
$Q^a_*{}_\alpha (\Pi_*\phi)_a=0$ which originates in the construction
of geodesic normal fields with the Vilkovisky connection. This is also
reflected in the fact that the Vilkovisky-DeWitt effective action
\begin{eqnarray}\label{VDeW} 
\Gamma_k[\bar\varphi]:=\Gamma_k[\bar\varphi,0] 
\end{eqnarray} 
is the standard background field effective action in the Landau-DeWitt
gauge, see e.g.\ \cite{Rebhan:1986wp}. Still it seems as if we have 
completely overcome the problem of solving modified Slavnov-Taylor 
identities. The flow \eq{flow} comprises the successive gauge
invariant integration of momentum fluctuations in $\Pi_* \bar\phi$ or
$\bar\varphi$ respectively. Moreover, it is accessible for practical
computations: one chooses a gauge invariant truncation of the
effective action and computes the flow of the scale dependent vertices
of $\Pi_*\bar\phi$ for fixed $\varphi_*=\varphi_0$, e.g.\ 
$\varphi_0=0$ or any other choice that is convenient. \smallstep 

However, the
effective action $\Gamma_k[\varphi_*,\bar\phi]$ only seemingly depends 
on two independent fields. It turns out that it depends on a
combination of these fields as derivatives w.r.t.\ the propagating
field $\bar\phi$ at fixed $\varphi_*$ are related to derivatives
w.r.t.\ $\varphi_*$ at fixed $\bar\phi$.  Part of the latter reflects
the base-point dependence of the construction constituting a gauge
dependence in the present formalism.  The corresponding relations are 
Nielsen identities, see e.g.\ \cite{Kunstatter:1991kw,Burgess:1987zi}. For the 
full effective action $\Gamma=\Gamma_0$ they read
\begin{eqnarray}
\Gamma_{,i}+
\Gamma_{,a}\langle \phi^a{}_{;i}\rangle=0\,, 
\label{genWI}
\end{eqnarray}
Using \eq{derivatives} the above equation can be rewritten in terms of 
$\hat\Gamma[\varphi_*,\bar\varphi]= 
\Gamma[\varphi_*,\bar\phi]$. 
\Eq{genWI} comprises the information that the $\varphi_*$-dependence of
$\Gamma,\hat\Gamma$ only enters via the source term
$(\phi-\bar\phi)\Gamma_{,a}$ in \eq{pathk} (at $k=0$). This is most
easily seen when rewriting the path integral in terms of an
integration over $\varphi$.\smallstep 

The gauge group structure is encoded in the dependence of $\phi$ on
the base point $\varphi_*$ and $\varphi$. The difference $(\langle
\phi^a{}_{;i}\rangle- \bar\phi^a{}_{;i})$ carries the information how
the classical realisation of the gauge symmetry is deformed within the
quantisation.  This fact is best seen in $\hat\Gamma_k$: without
quantum deformation of the gauge symmetry we had
$\hat\Gamma_{,i}^{(1,0)}=0$ as for the classical action, that is no
base point dependence. In the presence of the cut-off term the Nielsen
identities receive additional contributions. These terms can be
inferred from the path integral \eq{pathk} as $\langle\Delta
S_{k,i}\rangle$. They can be also derived in the spirit of the proofs
of {\bf (i), (ii), (iii)} in the previous section. This derivation is
 only based on the
validity of \eq{genWI} for the full effective action \cite{progress} without 
relying on the path integral representation \eq{pathk}. For
regulators satisfying \eq{onlyA} we arrive at
\begin{eqnarray}\label{genmNI}
\CW_{i}[\varphi_*,\bar\phi]&\equiv &0
\end{eqnarray} 
where
\begin{eqnarray*} 
\CW_{i}& :=& 
\Gamma_{k,i}-  
\s012 
G^{ab}\, R_{ba,i}+\bigl(
\Gamma_{k,a}-R_{ab}\, G^{bc}
\s0{\delta}{\delta \bar\phi^c}\bigr) \langle \phi^a{}_{;i}\rangle. 
\end{eqnarray*} 
\Eq{genmNI} can be rewritten in terms of
$\hat\Gamma_k[\varphi_*,\bar\varphi]= 
\Gamma_k[\varphi_*,\bar\phi]$ by using \eq{derivatives}. 
For vanishing cut-off \eq{genmNI} reduces to \eq{genWI}. \smallstep 

In the previous section we have learned that general regulators 
can be mapped to those satisfying \eq{onlyA}. Invariance under
transformations of $\varphi_*$, \eq{invariance2}, is lost. Instead we
arrive at modified Slavnov-Taylor identities for $\Gamma_k$ which
follows from \eq{genmNI} and \eq{invariance1}:
\begin{eqnarray}\label{mWIback}
Q_*^i{}_\alpha \Gamma_{k,i} = \s012 G^{ab} R_{ba,i}Q_*^i{}_\alpha. 
\end{eqnarray}
In \eq{mWIback} we have also used that
$Q_*^i{}_\alpha\phi^a{}_{;i}=Q_*^a{}_\alpha$, see \eq{bgaugetrafo},
and $\bQ_*^a{}_{\alpha,b}=0$. \Eq{mWIback} is similar to the
background field identities in the usual background field formalism
\cite{Freire:2000bq}. The structure is identical to that in the background
field approach to axial gauges \cite{Litim:2002ce}. In both cases 
the breaking of \eq{invariance2} is solely due to the regulator.  
The $t$-derivative
of \eq{mWIback} leads to \eq{step2}.  This is shown by using the flow
equation \eq{flow} for $\dot \Gamma_k^{(0,2)}$ in \eq{mWIback} and
\eq{mWIback} for $Q_*^i{}_\alpha \Gamma^{(0,2)}_{k,i}$ in \eq{step2}.
For regulators with \eq{Rproperty} the modified Slavnov-Taylor
identities \eq{mWIback} reduces to \eq{invariance2}. \smallstep

\Eq{genmNI} relates derivatives of the 
effective action and the expectation value $\langle \phi_{;i}\rangle$. 
Its flow is given by
\begin{eqnarray}
\partial_t \langle
\phi^a_{;i}\rangle =-\s012 (G\dot R G)^{bc}\langle
\phi^a{}_{;i}\rangle_{,cb}\,. 
\label{flow<>}\end{eqnarray} 
\Eq{flow<>} encodes the information how the non-trivial symmetry
structure of the theory at hand evolves with $k$.  Since $\langle
\phi_{;i}\rangle$ is not 1PI, it is a non-trivial check whether the
flow \eq{flow} of a truncated effective action $\Gamma_k$ stays a
solution of \eq{genmNI} if \eq{genmNI} is satisfied at some initial
scale $\Lambda$. The flow can be computed as
\begin{eqnarray}
\partial_t\CW_{i}& =& -\s012 
\bigl(G \dot R G\bigr)^{bc} \CW_{i,cb}\equiv 0.
\label{flowW}
\end{eqnarray}
 Thus
an approximation satisfying \eq{genmNI} at some initial scale
$\Lambda$ does so for all $k$. 
\Eq{flowW} has the same structure as the flow of modified
Slavnov-Taylor identities
\cite{D'Attanasio:1996jd,Litim:1998qi,Freire:2000bq,Litim:2002ce} which 
are 1PI identities.\smallstep

Finally we comment on the relevance of the Nielsen identities 
\eq{genmNI}. At first sight they pose no constraint at all and simply 
allow us to compute $\Gamma_k[\varphi_*,\bar\phi]$ from 
$\Gamma_k[\varphi_0,\bar\phi]$ in a Taylor expansion. 
The latter can be computed with 
the flow equation \eq{flow} from the initial effective action 
$\Gamma_\Lambda[\varphi_0,\bar\phi]$. \Eq{genmNI} can be evaluated
in orders of the fields as $\phi_{;i}$ has a covariant Taylor
expansion in $\phi$, the zeroth order term being $\delta^a{}_i$, see
e.g.\ \cite{Rebhan:1986wp}. Higher order terms depend on the
Vilkovisky connection \eq{Vcon} and (covariant) derivatives thereof.
It follows that $\langle \phi_{;i}\rangle[\varphi_0,\bar\phi]$ is a
function of $\Gamma_k^{(0,n)}[\varphi_0,\bar\phi]$, $
\varphi_0$ and $\bar\phi$. Hence with \eq{genmNI} we can
recursively determine $\Gamma_k^{(m,n)}[\varphi_0,\bar\phi]$ from
$\Gamma_k^{(0,n)}[\varphi_0,\bar\phi]$, in particular the Taylor
coefficients $\Gamma_k^{(m,0)}[\varphi_0,\bar\phi]$. Only
truncations $\Gamma_k[\varphi_0,\bar\phi]$ leading to a converging or
at least asymptotic series in $\varphi_*$ should be considered. This
already poses a non-trivial but unfortunately inaccessible constraint
on $\Gamma_k[\varphi_0,\bar\phi]$. \smallstep

There is a stronger and more relevant constraint on
$\Gamma_k[\varphi_0,\bar\phi]$. \Eq{genmNI} involves the projection
$\Pi_*$ and its derivatives (via the Vilkovisky connection) that in
general lead to infra-red singularities in the
Taylor coefficients $\Gamma_k^{(0,n)}[\varphi_0,\bar\phi]$ and
consequently in $\Gamma_k[\varphi_*,\bar\phi]$ for a completely
regular $\Gamma_k[\varphi_0,\bar\phi]$. Here, {\it regular} stands for
avoiding these singularities. Only truncations
$\Gamma_k[\varphi_0,\bar\phi]$ avoiding such pathologies should be
considered. This obstruction can be evaluated within a given
truncation as a fine tuning condition at the initial scale: regularity
of $\Gamma_\Lambda[\varphi_*,\phi]$ is equivalent to relations between
the $\Gamma_k^{(0,n)}[\varphi_0,\bar\phi]$.  An initial {\it regular}
effective action $\Gamma_\Lambda[\varphi_*,\phi]$ which satisfies
\eq{genmNI} (in some truncation), stays a {\it regular} solution at all
scales $k$, if the regulator is a sufficiently smooth function in
$\varphi_*$. This statement is proven by evaluating the flow
equation. If $\Gamma_\Lambda[\varphi_*,\phi]$ is regular, so is
$\Gamma_\Lambda^{(0,2)}[\varphi_*,\phi]$. As long as the regulator
does not introduce singularities, the flow is regular.  Hence,
regulators $R$ should be sufficiently smooth functions of $\varphi_*$
\cite{smooth}. We conclude that the situation here is comparable with
the fine tuning due to the modified Slavnov-Taylor identities in the
gauge fixed approach.

\section{Applications}

We illustrate the non-trivial content of the above constraints and the
non-local nature of the variables $\bar\phi$, $\bar\varphi$ within a
simple example. To begin with, we reproduce the standard one loop
effective action, see also \cite{Branchina:2003ek}. To that end we use
as input at the initial scale
\begin{eqnarray}\label{initialG}
\Gamma_\Lambda[\varphi_*,\bar\phi]=S+\s012 
(\ln \det \GG_{\alpha\beta}).
\end{eqnarray} 
see \eq{decompose}.  In \eq{initialG} we have used the fact that the
redundant fields $\phi^\alpha$ are not regularised, see \eq{onlyA}.
The related integration is trivial and removes the Fadeev-Popov
determinant $J$ \eq{J}. Within the one loop approximation the only
$t$-dependence on the right hand side of the flow \eq{flow} is the
explicit one in $R$. We are led to
\begin{eqnarray}
\Gamma_k^{\rm 1loop}[\varphi_*,\bar\phi]=S+\s012 
(\ln \det \GG_{\alpha\beta})
+
\s012 \left.(\Pi_*\ln(S^{(0,2)}[\varphi_*,\bar\phi]+R+P\kappa P]
)_a{}^a\right|_\Lambda^k. 
\label{1loop}\end{eqnarray} 
At $k=0$, $\Lambda\to\infty$ this result coincides with the one loop
geometrical effective action. 
For $\bar\phi=0$ this expression agrees
with the effective action obtained within the background field
formalism in the Landau-DeWitt gauge for vanishing fluctuation field.
Recursively one can show with \eq{flow} that these relations remain 
formally valid at higher loops.\smallstep

 The effective action in the Landau-DeWitt gauge satisfies 
Slavnov-Taylor identities whose non-trivial information is also
encoded in the Nielsen identities \eq{genmNI} in the geometrical
approach. Let us apply a simple pathological truncation to the initial
effective action \eq{initialG} in which the consequences of singular
solutions or even a breaking of \eq{genmNI} is evident. At the initial
scale we choose
\begin{eqnarray}\label{trunc}
\Gamma_\Lambda[\varphi_*,\bar\phi]=\s012 
\bar\phi^a S^{(0,2)}_{,ab}[0,0]\bar\phi^b+\varphi_*^i\,
\CF_i[\varphi_*,\bar\phi], 
\end{eqnarray}
where $\Gamma_\Lambda^{(0,n)}[0,\bar\phi]\equiv 0$ for $n>2$ and $\CF_i$ is
recursively determined by \eq{genmNI} in a Taylor
expansion about $\varphi_*=0$. A priori it is not even clear whether
such a procedure leads to an asymptotic series for
$\Gamma_\Lambda[\varphi_*,\bar\phi]$. Still \eq{trunc} is, if it
exists, gauge invariant \eq{invariance} and gauge independent: the
first term on the right hand side of \eq{trunc} satisfies
\eq{invariance1} with $\varphi_*=0$: 
$Q^a{}_\alpha[0]\, S^{(0,2)}_{,ab}[0,0]=0$ (transversal in plain momentum). 
As the remaining terms are determined by the
\eq{genmNI}, the statement follows.  Remarkably the truncation leads
to a trivial flow of $\Gamma_k[0,\bar\phi]$:
$\partial_t\Gamma_{k,a}[0,\bar\phi]=0$ and
$\Gamma_k[\varphi_*,\bar\phi]$ always takes the form \eq{trunc}.  It
looks highly non-trivial as it depends on arbitrary powers of
$\bar\phi$ or $\bar\varphi$ respectively. The same is true for the
Vilkovisky-DeWitt effective action $\Gamma_k[\bar\varphi]$ in
\eq{VDeW}. However, the trivial flow expresses the fact that we
completely lost the structure of the underlying non-Abelian gauge
symmetry genuinely leading to an interacting theory.  \smallstep

The above construction is impossible in a
standard gauge fixed approach as it violates the modified
Slavnov-Taylor identities: a kinetic term for the propagating fields
always leads to dynamics. By its nature, BRST symmetry is local and
the modification in the presence of the cut-off can be seen as a
perturbation. In the geometrical approach the apparent symmetry is
trivialised at the cost of non-localities \cite{costs}. The
non-trivial structure is still present and is carried by \eq{genmNI}.
The Taylor coefficients in the second term on the right hand side of
\eq{trunc}, recursively computed with \eq{genmNI} contain infra-red
singularities at all scales $k$, including $k=0$.  We have emphasised in the
beginning that it is not even clear whether
$\Gamma_k[\varphi_*,\bar\phi]$ exists at all for the present
truncation. As we have no closed form for
$\langle\bar\phi^a{}_{;i}\rangle$ it is difficult to decide. However,
\eq{trunc} with $\Gamma_\Lambda^{(0,n)}[0,\bar\phi]\equiv 0$ cannot
satisfy the 'classical' Nielsen identities
$\hat\Gamma^{(1,0)}[\varphi_*,\bar\varphi]=0$. Hence, there is at
least no solution of \eq{genmNI} with the form \eq{trunc} for the
effective action which admits a perturbative expansion in orders of
the coupling. Certainly, with this pathological example we have
artificially pushed the formalism beyond its limits, but it
pinpoints the relevance of fine-tuning infra-red regularity with
\eq{genmNI} as discussed in the previous section.  \smallstep

On the basis of the above findings we discuss possible applications 
of the approach as well as the relation of specific truncations to the
standard approach with modified Slavnov-Taylor identities.  The
regulator $R$ should provide a momentum cut-off. According to
\eq{Rproperty} or \eq{preRproperty} the corresponding momentum must be
gauge covariant. Options are e.g.\ $S^{(0,2)}[\varphi_*,0]$,
$\Gamma_k^{(0,2)}[\varphi_*,0]$ and combinations thereof. The
renormalisation group properties of such regulators have been
discussed in \cite{Pawlowski:2001df}. \smallstep

For
practical computations we have to commit ourselves to specific
coordinates which in most cases will be given by $\varphi_*^i$ or
$\bar\varphi^i$. A straightforward choice in Yang-Mills theory is
$\varphi_*=0$. At face value, the computation of
$\Gamma_k[0,\bar\phi]$ truncated in finite powers of $\bar\phi$ is
related to Landau gauge Yang-Mills, where, in first order, we have
substituted the ghost operator $\partial_\mu D_\mu$ by $D_\mu D_\mu$
(originating in $\GG$ in \eq{decompose}, see also \eq{initialG}).
Regularity computed from \eq{genmNI} puts a constraint on the momentum
structure of the propagator of $\bar\phi$.  Only at higher powers of
the field it also relates different vertices. It is in the spirit of a
truncation in finite powers of the fields to consider the latter
constraint as sub-leading.  When naively ignoring the remaining
differences we expect similar truncations to be relevant in Landau
gauge Yang-Mills. The latter indeed admits successful truncations that
relate the non-trivial momentum behaviour of propagators with
Slavnov-Taylor identities, by neglecting those leading to non-trivial
vertices, see e.g.\ \cite{Alkofer:2000wg}. \smallstep

Given the
subtleties concerning the physical content of truncations in
$\bar\phi$ and $\bar\varphi$ respectively it could be more promising
to identify the base point with the mean field:
$\varphi_*=\bar\varphi$, see also \cite{Branchina:2003ek}. The choice
of a physical mean field $\bar\varphi$ stabilises the flow. It
justifies truncations in powers of the fields as only fluctuations
about the physical mean should be important. This reduces the danger
of missing relevant physical information within a truncation as well
as that of caustics spoiling the construction
\cite{DeWitt:1998eq,Paris:en,Rebhan:1986wp}. For the same reason it
softens the effects of solving truncated Nielsen identities
\eq{genmNI}. On the practical side, it allows for the application of heat
kernel techniques which have already proven valuable within the
standard background field approach.  Such a procedure is close to
using the Landau-DeWitt gauge in the background field formalism and
similar truncations to the modified Slavnov-Taylor identities. This
supports the gauge invariant truncation schemes used in the background
field approach. \smallstep

As discussed in the introduction, in the background field approach we
do not have full access to the modified Slavnov-Taylor identities for
gauge invariant truncations. In the geometrical approach we have full
access to the symmetry constraints \eq{genmNI} and gauge invariance is
guaranteed by the construction. This is the the real benefit of the
present formalism. Still, \eq{genmNI} poses a considerable challenge
even within a truncation. It contains a loop term related to the
regulator and its $\varphi_*$-derivatives as well as $ \langle
\phi^a_{;i}\rangle$ which cannot be presented in a closed
form. \smallstep

A promising practical strategy for an approximate solution of
\eq{genmNI} is based on the flow equation \eq{flow} itself: given a
sufficiently smooth regulator $R$, we consider truncations $\Gamma_k$
admitting a regular $\Gamma_k^{(0,2)}[\varphi_*,\bar\phi]$.  For the
sake of simplicity, let us consider a truncation relying on only a
finite number of powers in $\varphi_*,\bar\phi$ or
$\varphi_*,\bar\varphi$ respectively and a fixed momentum behaviour of
the vertices. This amounts to assuming 
\begin{eqnarray}\label{truncelab} 
\Gamma_\Lambda^{(n,m)}\equiv 0 &\quad {\rm for} \quad & 
n> N_0,\quad  m> M_0 
\end{eqnarray}
and demanding infra-red regular initial vertices
\begin{eqnarray}\nonumber 
&&\Gamma_k^{(n,m)}[0,0](p_1,...,p_{n+m}), \quad   
R^{(n,0)}(p_1,...,p_{n})\quad {\rm for} \quad 
n\leq N_0,\quad  m\leq M_0
\label{demand} \end{eqnarray}
with fixed momentum dependence. As we have shown in section~4, even
the most general regulator is effectively $\varphi_*$-dependent and
hence $R^{(n,0)}\neq 0$ for at least some $n$, see \eq{effective}.
Regularity of \eq{demand} guarantees a regular flow for
$\Gamma_k^{(n,m)}$ with $n\leq N_0,\,m\leq M_0$. Truncating also
$\langle \phi^a{}_{;i}\rangle$ at a finite power in the fields leads
to a closed expression of \eq{genmNI}. The relative coefficients of
the initial vertices $\Gamma_\Lambda^{(n,m)}$ are fixed by \eq{genmNI}
which also depend on $R^{(n,0)}(p_1,...,p_{n})$. The fact that $R$ is
up to our disposal (within the above mentioned restrictions) might be
used to facilitate the above task. For example one could demand
partial cancellation of $\Gamma_k^{(n,m)}$ and $R^{(n,0)}$ terms.
Another observation is that for a sharp cut-off $R\,G=0$ and the last
term in $\CW_{k,i}$ \eq{genmNI} vanishes.  Note that this comes to the
price of complicating the computation of $G\, R_{,i}$-contributions.
The strategy described above can be quite generally applied to other
truncation schemes. It guarantees that the Vilkovisky-DeWitt effective
$\Gamma_k[\bar\varphi]$ action \eq{VDeW} is infra-red regular during
the flow. Physical observables are directly accessible via the vertices 
of $\Gamma_k[\bar\varphi]$. 
Results obtained in gauge theories and
Euclidean quantum gravity will be published elsewhere.\smallstep

The approach presented here is also of interest for the 2PI formalism
\cite{Cornwall:vz} which has recently experienced renewed interest. 
It hinges on
coupling an external source to the propagator $G$ and performing a
second Legendre transform. There is a close connection to the flow
equation as the regulator $R$ can be interpreted as such a source.
Flow equations for 2PI fermionic systems have recently been studied in
\cite{Wetterich:2002ky}. In gauge theories both formalisms have to
fight similar problems as the propagator of the gauge field is not a
gauge invariant object. These problems have been partially resolved
within a background field approach to the 2PI effective action, see
e.g.\ \cite{Mottola:2003vx}, where on-shell gauge invariance has been
proven. Still, as in the background field approach to flow equations,
one deals with non-algebraic Slavnov-Taylor and Nielsen identities. It
seems to be natural to apply the present ideas within the 2PI approach
gaining off-shell gauge invariance, effectively reducing the theory to
a scalar one. Then, even without regulator one deals with Nielsen
identities similar to \eq{genmNI} \cite{progress}. As the regulator adds no
further difficulty, one may as well use the flow equation on top of
the 2PI formulation for its
numerical merits.

\section{Summary}

We have derived a flow equation for the geometrical
effective action. For the sake of simplicity we have restricted
ourselves to non-Abelian gauge theories in flat space-time but flows
for quantum gravity and Yang-Mills in curved space-time are
straightforward extensions of the flow presented here. \smallstep

The flow \eq{flow} is gauge invariant, gauge 
independent and allows for general cut-off functions. The latter is
important for practical applications, where the optimisation of the
flow relies on the possibility of freely varying the regulator. The
present construction is genuinely non-local, both in momentum as well
as in field space \cite{costs}. This makes the generality of the
regulator even more desirable, as the cut-off has to be sufficiently
smooth \cite{smooth}.  We expect that physically interesting
truncation schemes include a non-trivial momentum behaviour for
propagators and vertices. This must be reflected in the choice of an
appropriate regulator.  \smallstep 

It has been shown that gauge symmetry is realised rather non-trivially
in the modified Nielsen identities \eq{genmNI}, despite gauge
invariance of the construction. These constraints are closely linked
to the modified Slavnov-Taylor identities. Their solution guarantees
the absence of pathological infra-red singularities. In given
truncations they can be simplified by using the freedom in the
regulator. In contradistinction to the background
field approach we have full access to the symmetry constraints in
gauge invariant truncations. This allows for a self-contained
systematic approach. In comparison to general gauge fixed formulations
we gain gauge invariance leading to a direct access to observables. 
\smallstep 

We have also discussed how practical computations 
can be put forward in the formalism as well as their relation to
computations within the standard background field approach and those
in Lorentz-Landau gauge.  In short, the results here back-up the
standard truncations already used in the gauge fixed approach as
well as providing a new, promising avenue to go beyond. We also briefly 
outlined how to use the present ideas in a 2PI approach.\step 

\acknowledgements 

This work owes much to insights gained within unpublished 
work with C.
Mol\'{\i}na-Paris. I also thank D.~O'Connor and P.~Watts for discussions.


\begin{thebibliography}{99}
\bibitem{strictly} 
For $R\to \infty$ the
  proofs require an additional perturbative recursion. For $R\to 0$ 
this is not necessary.  Still 
some care is required as typically $\partial_k R|_{k=0}\equiv 0$. Strictly
speaking the proofs with $\Gamma_0$ apply to the (hull of the) set
of regulators with $\partial^n_k R|_{k=0}\neq 0$ for some minimal $n_0$ 
and using $\partial^{n_0}_k \Gamma_k|_{k=0}\neq 0$.
\bibitem{smooth} Sufficient smoothness is a truncation scheme
  dependent statement. Even though differentiability of $R$ certainly
  facilitates the task it is, a priori, not necessary. A simple
  example is provided by the sharp cut-off in a scalar theory and the
  derivative expansion about $p^2=0$. This truncation scheme is not
  working properly at higher orders but is fine for expanding about
  more suitable expansion points
\bibitem{costs} As such, non-localities pose no problem within the
  flow equation. This is actually one of its merits
 and is well suited for the geometrical approach.
\end{thebibliography}
\end{document}